\documentclass{iopconfser}

\usepackage{graphicx} 
\usepackage{subcaption}

\begin{document}

\title{Investigating all-sky Frequency Hough performances for neutron stars}

\author{Martina Di Cesare$^{1,2}$, Pia Astone$^{3}$, Rosario De Rosa$^{1,2}$, David Keitel$^{5}$, Cristiano Palomba$^{3}$, Marco Serra$^{3}$}

\affil{$^1$ Università degli Studi di Napoli "Federico II", I-80126 Napoli, Italy
\\$^2$ INFN, Sezione di Napoli, I-80126 Napoli, Italy
\\$^3$ INFN, Sezione di Roma, I-00185 Roma, Italy
\\$^4$ Università di Roma “La Sapienza", I-00185 Roma, Italy
\\$^5$ Departament de Física, Universitat de les Illes Balears, IAC3–IEEC, Crta. Valldemossa km 7.5, E-07122 Palma, Spain}

\email{martina.dicesare@unina.it}

\begin{abstract}
Between the estimated population of Neutron Stars (NSs) and the actual number present in the catalogs, there is a huge gap: O(10$^{8-9}$) vs O(10$^3$). Among the different search techniques for Continuous gravitational waves (CWs), the all-sky could help to reduce the discrepancy. We focus on the all-sky CW pipeline Frequency Hough (FH), which operates without prior knowledge of the source parameters ($f,\dot{f}, \lambda, \beta$). Here, we present a Machine Learning strategy, diverging from the standard follow-up(FU) of the FH pipeline. We study the performance with real interferometer data, until reaching $h$ value subthreshold for the standard FU procedure ($CR_{thr}=5$), with encouraging classification results.
\end{abstract}

\section{Continuous waves approaches}
The Gravitational Wave (GW) signal from a periodic source, such as a neutron star (NS), is called continuous, producing small metric perturbations over time. This motivates integrating data to extract the continuous wave (CW) signal from noise. An isolated NS is described by ${f,\dot{f},\lambda,\beta}$, which are frequency, spin-down, longitude, and latitude, respectively. Search strategies, in order of decreasing sensitivity and increasing computational cost, are: targeted, narrow-band, directed, and all-sky — the last being the most expensive due to the lack of prior information. The aim is to reduce the gap between the predicted NS population ($\rm O(10^{8-9})$) and those observed ($\rm O(10^3)$) in the Milky Way \cite{Reed_2021}.

\section{Frequency Hough pipeline}
\begin{figure}[ht]
    \centering
        \centering
        \includegraphics[width=0.8\linewidth]{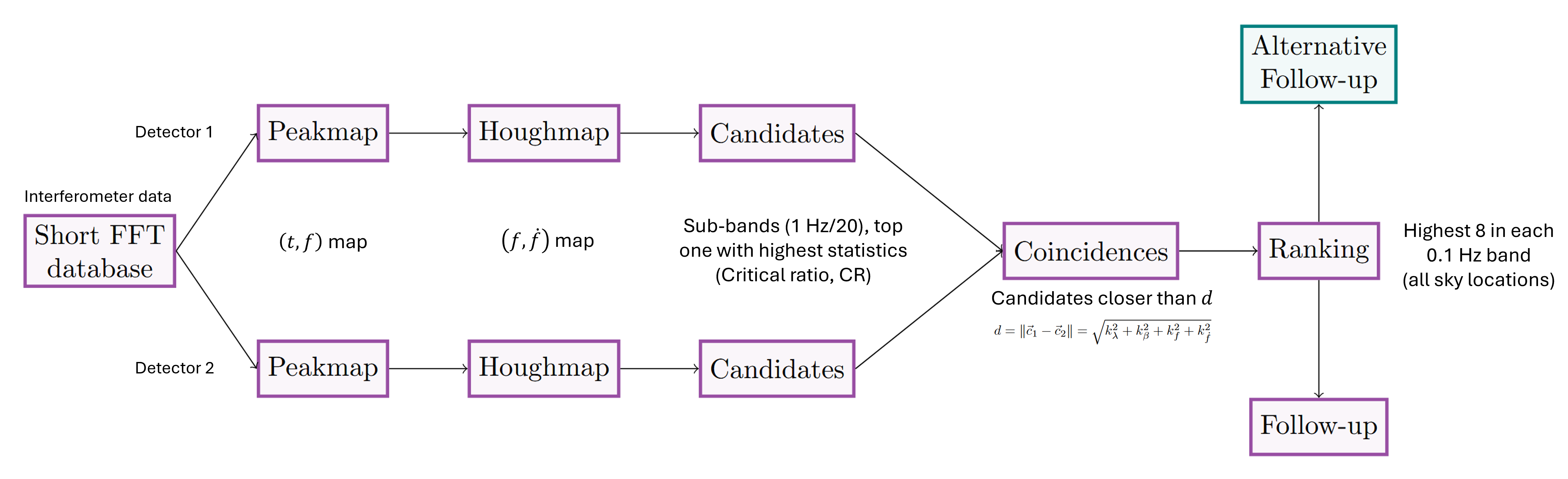}
        \caption{Frequency Hough pipeline standard workflow (purple), with the Machine Learning approach (green).}
    \label{fig:FH_workflow_labels}
\end{figure}

The Frequency Hough (FH) pipeline is an all-sky method to search for CW from isolated NS \cite{PhysRevD.90.042002}. In the third observing run (O3), the workflow is the one presented in Fig.~\ref{fig:FH_workflow_labels}. In particular, a CW signal’s strength is quantified by 
$n=\sum_{i}^{N_{FFT}} n_{i}w_{i}$ (adaptive); the proposed alternative approach version uses 
$n=\sum_{i}^{N_{FFT}} n_{i}$ (non-adaptive, more noise-sensitive). The $n$ is normalized with respect to the $\mu,\sigma$ of the Houghmap (HM), returning the Critical Ratio, i.e., $CR=\frac{n-\mu}{\sigma}$. A special focus is given to the follow-up, since this is the diverging step from the standard analysis. Follow-up takes candidates $CR\ge 5$, repeating the FH steps with $T_{FFT}^{FU}=3\cdot T_{FFT}$ and heterodyne Doppler correction for $f^{cand},\dot{f}^{cand}$ and nearby sky points, seeking CR growth.

\vspace{-15pt}

\section{Alternative follow-up with Neural Network (NN)}
\begin{figure}[ht]
    \centering
    \includegraphics[width=0.8\linewidth]{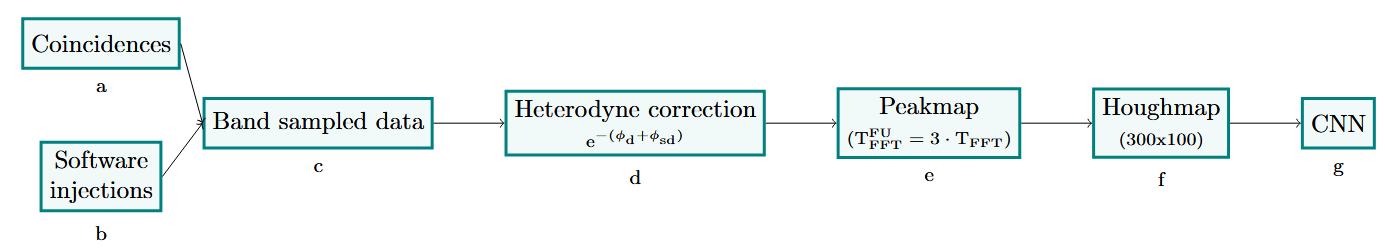}
    \caption{Alternative FU workflow.}
    \label{fig:Alternative_FU_workflow}
\end{figure}

We investigate the performance of an alternative follow-up (Fig.~\ref{fig:FH_workflow_labels}, green box) regarding its accuracy and speed on tagging correctly an HM as containing a potential GW signal or being just a noise map, allowing the analysis of candidates below $CR=5$. The workflow starts by selecting from the O3 Short FastFourier database (SFDB) 1 Hz, thanks to the Band Sampled Data (BSD) framework \cite{Piccinni_2019}. The 1 Hz step is the narrowest bandwidth allowed to avoid generating artifacts with the BSD procedure. This procedure is done for candidates surviving the first-stage selection, where we build an HM around the
candidate’s parameters after demodulating with a heterodyne phase using ($f_{cand}, \dot{f}_{cand}, \lambda_{cand},\beta_{cand}$). In particular, the resolution is set by the enhanced coherence time $T_{FFT}^{FU}=3\cdot T_{FFT}$, Fig. \ref{fig:Alternative_FU_workflow}. Moreover, a skygrid $(\lambda_{i},\beta_{i}$) is built around the candidate's location, using the sky position for the heterodyne Doppler correction. This is a further test to investigate the model's ability to recognize a signal with respect to its true sky coordinates.

We also keep track of the execution times required for building the maps to check the feasibility of the approach. These are, following the labels in Fig. \ref{fig:Alternative_FU_workflow} for $e$: 16.07 s; $f$: 2.24 s; $e\rightarrow f$: 64.58 s on NVIDIA L40S.

\subsection{Collection of HM}
\begin{figure}[ht]
    \centering
    \includegraphics[width=0.85\linewidth]{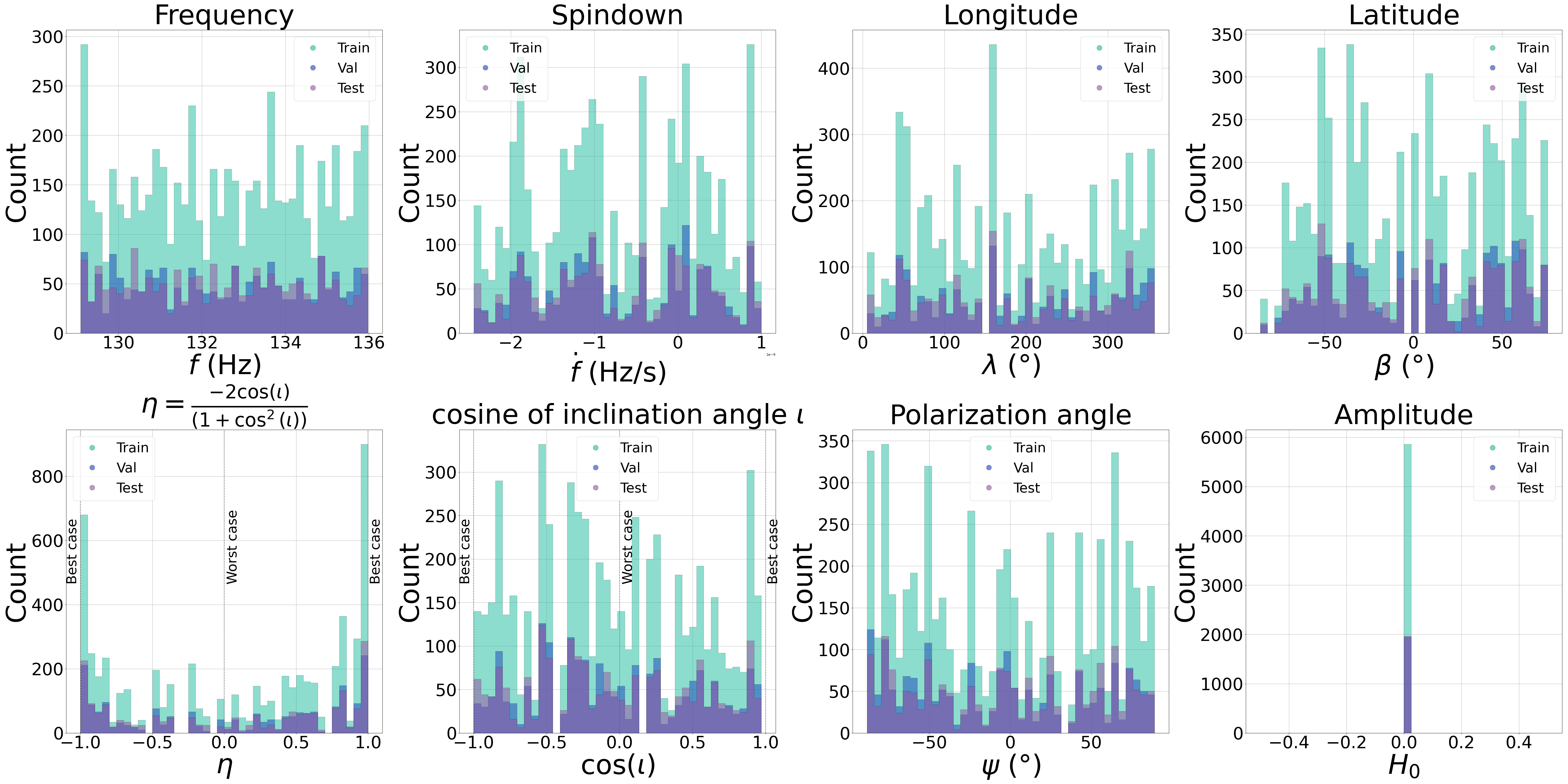}
    \caption{Repartition of the HM for the three stages (Train, Validation, and Test).}
    \label{fig:histogram}
\end{figure}

To ensure unbiased training, we take care of selecting an equal number of Signal and Noise maps, with an extra focus on the parameters of the Signal HMs, which have to span across the parameter space. The combined Signal-Noise set is then divided into Train (60\%), Validation (20\%), and Test (20\%) subsets. Train minimizes the loss by adjusting kernel weights; Validation checks performance; Test uses unseen data. Fig.~\ref{fig:histogram} shows sample subsets. Noise maps come from O3 candidates, which have been investigated with the standard FH and discarded as a GW source; this implies that they were generated by noise in the data. Signal maps from software injections (SIs) with fixed amplitude, while $\cos\iota \in [-1,1],\psi \in [-90^{\circ}, 90^{\circ}]$ follow a uniform distribution. SIs are added to real O3 data, and HMs are normalized per detector by FFT count, using the non-adaptive case.

\newpage

\subsection{Results}
\begin{figure}[ht]
    \centering
    \begin{subfigure}[b]{0.4\linewidth}
        \centering
        \includegraphics[width=\linewidth]{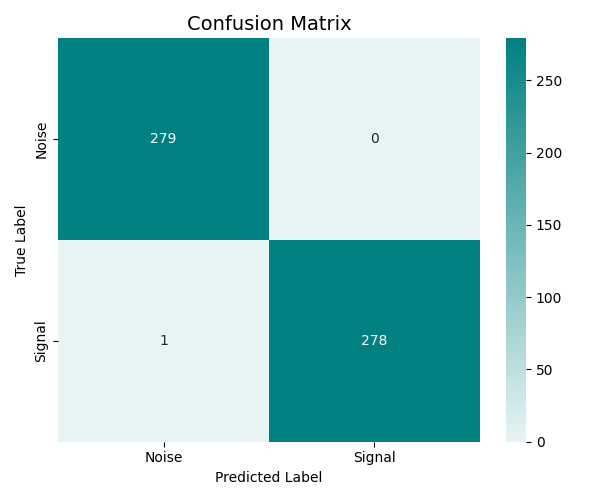}
        \caption{LIGO Livingston.}
        \label{fig:conf_LL}
    \end{subfigure}
    \hfill
    \begin{subfigure}{0.4\textwidth}
        \centering
        \includegraphics[width=\linewidth]{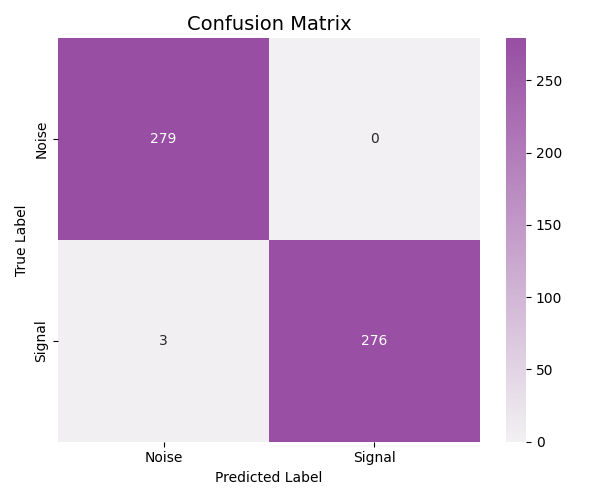}
        \caption{LIGO Hanford.}
        \label{fig:conf_LH}
    \end{subfigure}
    \caption{Test results of the two LIGO detectors displayed with a confusion matrix.}
    \label{fig:conf_LL_LH}
\end{figure}

Studies with decreasing values of $H_{0}$ (GW amplitude used in the FH pipeline) were performed, showing an efficient classification of the maps to the right class. $H_{0}$ can be converted to $h_{0}$ by averaging the factor depending on $\cos\iota$, Eq. 67 \cite{PhysRevD.90.042002}, i.e. $H_{0}=h_{0}\cdot(1.32)$. The first tool to use for describing the performance of the model is looking at the Train and Validation Loss/Accuracy curves obtained during the analysis: they show that the model can learn the Noise and Signal patterns just after a few epochs. Instead, in the Test phase, we can evaluate it through the Confusion matrix. Here, we display the weakest amplitude so far, i.e., $H_{0}=1.02\cdot10^{-25}$. The main diagonal of the confusion matrix for the two LIGO detectors, Fig. \ref{fig:conf_LL_LH}, shows the HMs tagged correctly by the model in the third stage. The elements off the diagonal are the incorrect predictions. In the example provided, it can be appreciated the model's performance is even with a small dataset: almost the entire collection of HMs belongs to the main diagonal. In general, since for one SIs there are more HMs due to the skygrid, even just one correctly tagged HMs could allow us to catch the signal.

\vspace{-14pt}

\vspace{-28pt}

\section{Prospects and Conclusions}
The study introduces an NN-based alternative follow-up for the FH pipeline, exploring the computational cost and performance of labeling the HMs. We used real O3 data, decreasing the strain of the artificial signals injected. It has been shown, even at the smallest strain explored so far, that the model performs with an almost perfect score. Future plans include expanding datasets, adding binary signals, weaker signals, and optimizing training to improve generalization and map-based sky analysis.

\vspace{-25pt}

\section{Acknowledgments}
This material is based upon work supported by NSF's LIGO Laboratory, which is a major facility fully funded by the National Science Foundation.

\bibliographystyle{unsrt}
\bibliography{biblio}

\end{document}